\input{aipcheck}

\documentclass[
    ,final            
  ]
  {aipproc}

\layoutstyle{6x9}
\usepackage[rightcaption]{sidecap}

\begin{document}

\title{A Metal-biased Planet Search}

\classification{PACS: 97.82.Cp}
\keywords      {stars: planetary systems, stars: activity, stars: abundances}

\author{James S. Jenkins}{
  address={Universidad de Chile, Camino el Observatorio 1515, Las Condes, Santiago, Chile, Casilla 36-D}
}

\author{Hugh R.A. Jones, John R. Barnes, Yakiv Pavlenko}{
  address={Centre for Astrophysics, University of Hertfordshire, College Lane, Hatfield, Hertfordshire, AL109AB, UK}
}

\author{Patricio Rojo, Matias I. Jones}{
  address={Universidad de Chile, Camino el Observatorio 1515, Las Condes, Santiago, Chile, Casilla 36-D}
}

\author{Avril C. Day-Jones}{
  address={Centre for Astrophysics, University of Hertfordshire, College Lane, Hatfield, Hertfordshire, AL109AB, UK}
}

\author{David J. Pinfield}{
  address={Centre for Astrophysics, University of Hertfordshire, College Lane, Hatfield, Hertfordshire, AL109AB, UK}
}

\begin{abstract}

We have begun a metal-rich planet search project using the HARPS instrument in La Silla, Chile to target planets with a high potential to transit their 
host star and add to the number of bright benchmark transiting planets.  The sample currently consists of 100, bright 
(7.5~$\le$~$V$~$\le$~9.5) solar-type stars (0.5~$\le$~$B-V$~$\le$~0.9) in the southern hemisphere which are both inactive (log$R'$$_{\rm{HK}}$~$\le$~-4.5) 
and metal-rich ([Fe/H]~$\ge$~0.1~dex).  We determined the chromospheric activity and metallicity status of our sample using high resolution FEROS 
spectra.  We also introduce the first result from our HARPS planet search and show that the radial-velocity amplitude of this star is consistent with 
an orbiting planetary-mass companion (i.e. Msin$i$~$<$~0.5M$_{\rm{J}}$) with a period of $\sim$5 days.  We are currently engaged in follow-up to 
confirm this signal as a bonafide orbiting planet.

\end{abstract}

\maketitle


\section{Introduction}

Since we now have a compilation of over 300 extrasolar planets (aka. exoplanets) a number of trends and correlations have been found in the data.  For 
instance, \citet{fischer05} have recently confirmed the overabundance of metallicity for planet host stars compared with a comparison field star 
sample using a single homogenious technique to extract the metallicity values.  \citet{bond08} have shown that all heavy elements they measured for 
planet hosts on the Anglo-Australian Planet Search (AAPS) were overabundant compared to stars with no planets.  \citet{israelian04} and \citet{gonzalez08} have 
shown that there is an overdepletion of lithium in planet hosts compared with a field sample.  Such trends are extremely important in understanding the 
underlying formation/evolution mechanisms of exoplanets, however they are also great tools to use in the hunt for exoplanets around stars as yet 
unobserved.

\section{Activities and Metallicities}

\subsection{Activities}

\begin{SCfigure}
  \includegraphics[height=.33\textheight,angle=90]{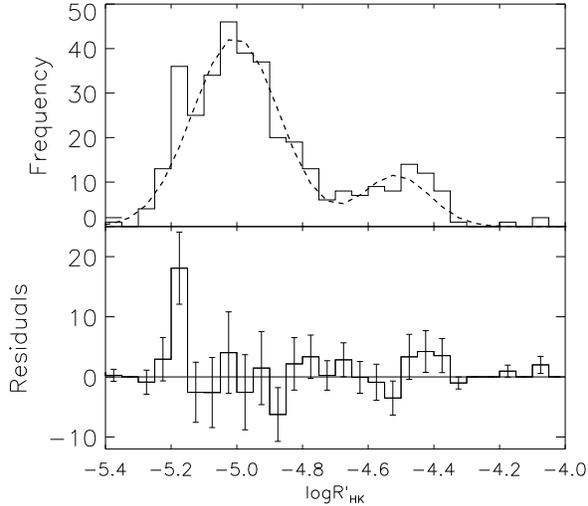}
  \caption{The upper plot shows the distribution of chromospheric activities (log$R'$$_{\rm{HK}}$) for our sample.  The dashed line is a best fit double 
gaussian to the data.  The lower panel shows the residuals to the fit along with their uncertainties and all but one of the bins agree with the fit 
to within $\pm$1$\sigma$.}
\end{SCfigure}

Radial-velocity precision now reaches down below the 3m/s limit (e.g. \cite{pepe03}; \cite{tinney06}; \cite{wright08}) so we are 
sensitive to the various motions of the stellar envelope.  In particular cool star spots traversing the stellar photospheres can cause erroneous scatter 
in the radial-velocity timeseries (\cite{saar01}; \cite{wright05}) or even induce false planetary signatures in the data (e.g. \cite{queloz}; 
\cite{henry02}) if the timeseries is sufficiently coupled to the stellar rotation.  Therefore, knowledge of the level of chromospheric activity can 
lead to a more targeted search for exoplanets as one can target the most quiescent stars to extract the most precise signals.  We extract the 
log$R'$$_{\rm{HK}}$ activity index from our high resolution FEROS sample and target the most inactive stars (log$R'$$_{\rm{HK}}$~$\le$~-4.5), 
ensuring we can reach the highest level of precisions.  Fig.~1 (upper) shows the final distribution for our sample of over 350 stars and we confirm the 
bimodal nature of stellar activities.  Also the majority of our sample are located comfortably above our planet search sample cutoff of -4.50 and should 
represent good quiescent radial-velocity targets.  The dashed line represents the best double gaussian fit to the data and has means of -5.00 and -4.52 
with variances of 0.10 and 0.13 respectively.  The lower panel shows the residuals to the fit with their associated uncertainties and it is found that all 
but one of bins are located within $\pm$1$\sigma$ of the fit.  This bin mostly contains stars evolving off the main sequence and since 
the log$R'$$_{\rm{HK}}$ index was formulated for stars on the main sequence (see \cite{noyes}) it may be gravity dependent and not fully applicable to 
such evolved stars.
 
\subsection{Metallicities}

\begin{SCfigure}
  \includegraphics[height=.34\textheight,angle=90]{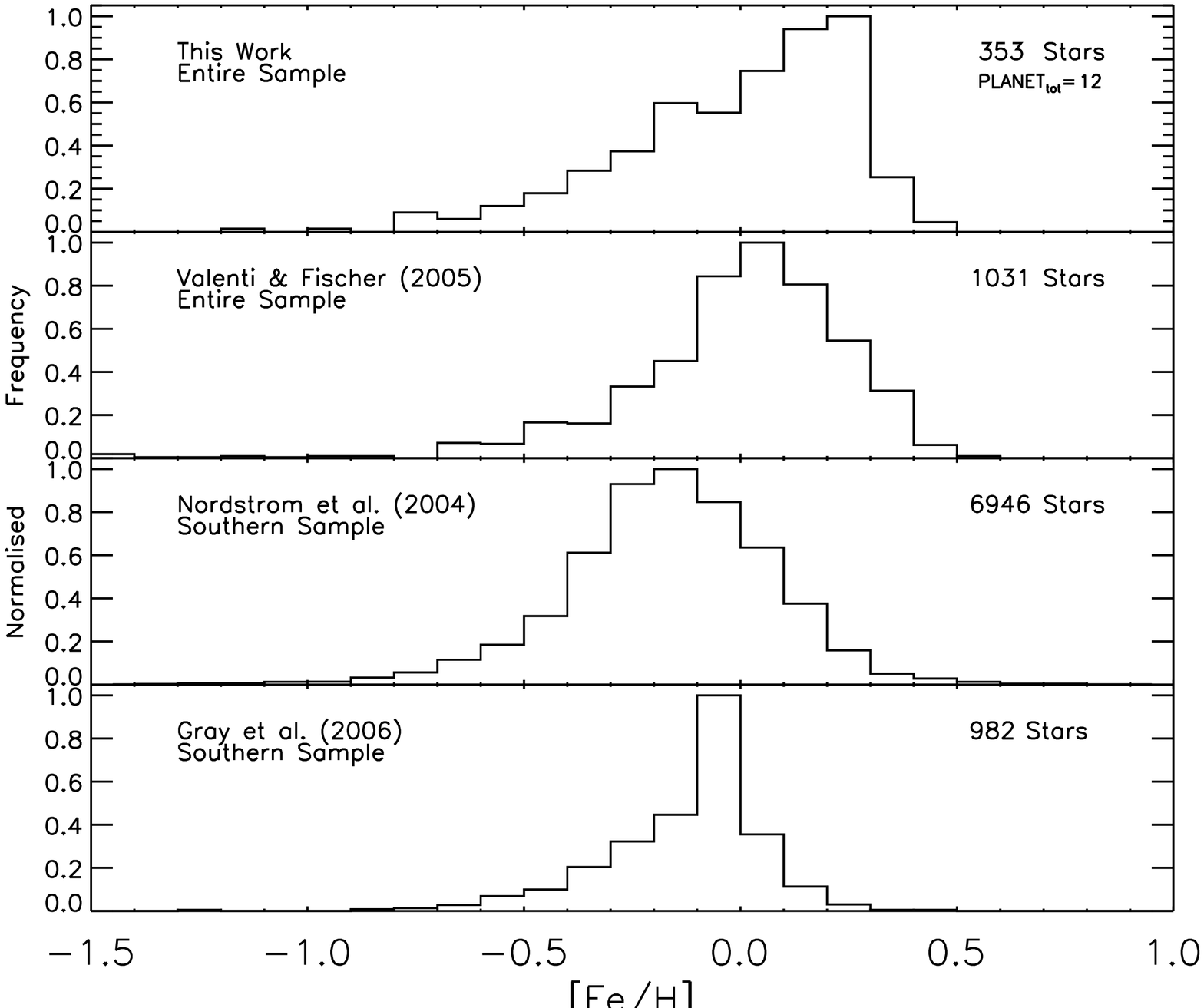}
  \caption{Four plots showing the distributions of metallicities of our sample (top) compared with three other works; \citet{valenti05} (upper middle), 
\citet{nordstrom04} (lower middle) and \citet{gray06} (bottom).  The top two panels are from planet search samples, whereas the bottom two are more 
general nearby star samples.  Also shown are the number of stars in each sample, along with the expected number 
of planets with signals $>$30m/s.  Note, we have considerably more potential planets since we have included signals of lower amplitudes.}
\end{SCfigure}

As mentioned above, the early discovery of the metal-rich nature of planet hosts (\cite{gonzalez}), which is described in the framework of the core accretion 
scenario of planet formation (e.g. \cite{pollack}; \cite{ida08}), has made us rethink the way stars are selected for 
planet search projects.  We measured the metallicity ([Fe/H]) of each of our target candidates and then extracted a subset of these which have 
[Fe/H]~$\ge$~0.1~dex for our HARPS planet search project.  The final distribution of metallicities in our sample is shown in the upper plot of 
Fig.~2, and it is clear that the distribution peaks in the metal-rich regime.  This is due to a bias introduced in our initial sample selection where some 
stars with photometric metallicities in the extreme metal-rich domain were added to the FEROS sample to help increase the number of 
spectroscopically determined metal-rich stars and helping to give rise to the current sample of 100 planet search targets.  The other three 
plots represent distributions from other works (\cite{valenti05} (upper middle); 
\cite{nordstrom04} (lower middle); \cite{gray06} (bottom)).  The Valenti \& Fischer sample peak in the metal-rich regime also, again due to a 
metal-rich bias in their sample since they included all stars on the AAPS, Keck and Lick planet searches.  However, both the 
lower two plots from Nordstr{\"o}m et al. and Gray et al. peak in the metal-poor regime, with the Nordstr{\"o}m et al. sample peaking at the most 
metal-poor value.  Given that they had the most kinematically unbiassed sample this should better represent the metallicity distribution of the 
galaxy for solar-type stars and highlights why we included a number of photometrically metal-rich stars to help bias our distribution into the 
metal-rich regime and increase our target sample.

\section{Current HARPS Results}

We began observations on HARPS in late 2006 and to date we have had 20 nights of time to compile a baseline of velocities, weed out spectroscopic 
binary signatures and target potential planetary signals in the data.  We follow the methodology of \citet{fischer05c} by observing three data points 
for each star and then densely sampling any potential planetary induced signatures.  Over our 20 nights of observations we have obtained three data 
points for almost all of our sample.  Currently, 30\% of our sample show candidate planetary signatures, with nearly half of these appearing to have short period 
orbits.  

\begin{SCfigure}
  \includegraphics[height=0.25\textheight]{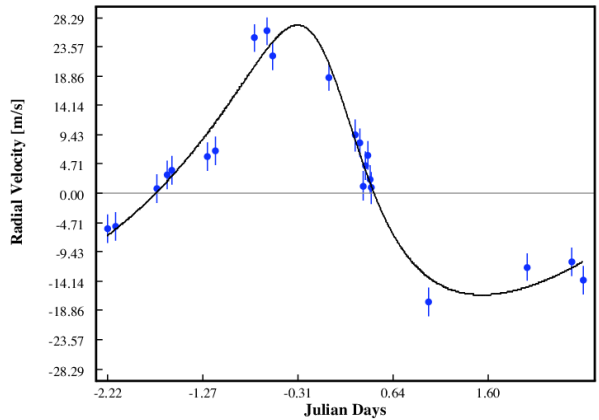}
  \caption{The phase folded Keplerian fit to the exoplanet candidate host.  The Keplerian has a period of $\sim$5~days and has moderate 
eccentricity.  Across the two epochs we have mapped the entire orbit, twice sampling periastron passage.}
\end{SCfigure}

Fig.~3 shows an example phase folded Keplerian curve for one of our target stars using Systemic (\url{http://oklo.org}).  It is 
clear that the 
important periastron passage has been well constrained.  The 
eccentric signal was reproduced in both epochs, split by over 4 months, showing it has been coherent for at least this time, arguing against 
the activity hypothesis.  However, we are currently engaged in follow-up of this object to confirm that the signal is due to 
an orbiting planet and not due to any photospheric activity cycles like star spots.



\bibliographystyle{aipproc}   

\bibliography{refs}

\IfFileExists{\jobname.bbl}{}
 {\typeout{}
  \typeout{******************************************}
  \typeout{** Please run "bibtex \jobname" to optain}
  \typeout{** the bibliography and then re-run LaTeX}
  \typeout{** twice to fix the references!}
  \typeout{******************************************}
  \typeout{}
 }

\end{document}